\documentclass[apjl]{emulateapj}
\usepackage{amsmath,amssymb,graphicx,soul,color}
\usepackage{epstopdf}
\usepackage{subfigure}

\shorttitle{MHD oscillations in flaring stellar loops}
\shortauthors{Pandey \& Srivastava}

\begin{document}

\title{Observations of X-ray oscillations in $\xi$ Boo:  Evidence of a fast-kink mode in the stellar loops}
\author{J. C. Pandey and A. K. Srivastava}
\affil{Aryabhatta Research Institute of Observational Sciences, Naini tal-263 129, India}

\begin{abstract}
We report the observations of X-ray oscillations during the flare in a cool active star $\xi$ Boo for the first time.
$\xi$ Boo was observed by EPIC/MOS of {\it XMM-Newton} satellite. The X-ray light curve is investigated with wavelet and periodogram analyses. Both analyses clearly show oscillations of the period of $\sim 1019$ s. We interpret these oscillations as a fundamental fast-kink mode of magnetoacoustic waves.

\end{abstract}


\section{Introduction}
The  magnetohydrodynamic (MHD) waves and oscillations in the solar
and  probably in the stellar atmospheres are assumed to be generated by coupling of complex magnetic field and plasma.  These MHD waves and oscillations are one of the
important candidates for heating  the solar/stellar coronae and accelerating the supersonic winds. In the Sun, magnetically structured flaring  loops,
anchored into the photosphere, exhibit various kinds of MHD
oscillations (e.g. fast sausage, kink and slow acoustic oscillations).
The idea of exploiting such observed oscillations as a diagnostic tool for deducing
physical parameters of the coronal plasma was first suggested by Roberts et al. (1984) . Wave and oscillatory activity of the solar corona is now well observed with modern imaging and spectral instruments (e.g. instruments on-board {/it SOHO, TRACE, HINODE and NoRH} ) in the visible, EUV, X-ray and radio bands, and interpreted in terms of MHD wave theory (Nakariakov \& Verwichte 2005).

Stellar MHD seismology is also a new developing tool to deduce the physical properties
of the atmosphere of magnetically active stars, and it is based on the analogy
of the solar coronal seismology (Nakariakov  2007). The first stellar X-ray flare oscillations in
AT Mic have been reported by Mitra-Kraev et al. (2005) and it was interpreted as a signature of slow
magneto-acoustic oscillation.  Mathioudakis et al. (2006) have reported the fast sausage
oscillations of $\sim 10$ s in the flaring stars  EQ Peg B. Mullan et al. (1992) have also
found the oscillations on the scale of few minutes and interpreted as a transient oscillations
in the stellar loops. In this Letter, we find the observational signature of the fast kink oscillations in the magnetically active star $\xi$ Boo (V = 4.55 mag). In the case of solar corona, such type of modes are important to probe the local plasma conditions (e.g., van Doorsselaere et al. 2008). Therefore, such modes may also  be useful to diagnose the stellar coronae.  $\xi$ Boo is a nearby (6.0 pc) visual binary, comprising a primary G8 dwarf and a secondary K4 dwarf with an orbital period of 151 yr. In terms of the outer atmospheric emission, UV and X-ray observations show that the primary dominates entirely over the secondary (e.g. Laming \& Drake 1999). The observations, data reduction, results and discussion  along with the  conclusions are given in the forthcoming sections.

\section{Observations and data reduction \label{sec:obs}}
\begin{figure}
\centering
\includegraphics[width=50mm,angle=-90]{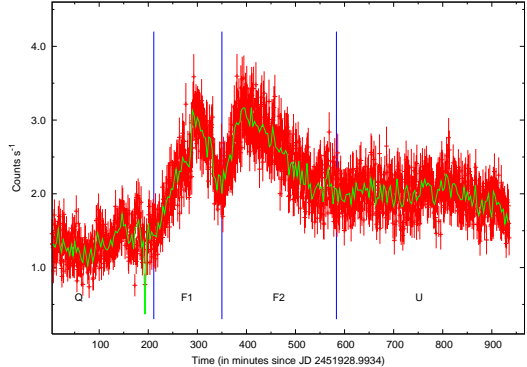}
\caption{The X-ray 40-s binned light curve of the star $\xi$ Boo in the energy band 0.3-10. keV. The overplotted continuous line is 200 s binned light curve}
\label{fig:lc}
\end{figure}

The star $\xi$ Boo was observed  by RGS  and EPIC-MOS  on boards XMM-Newton satellite (Jansen et al. 2001) on 2001 January 19 at 19:25:06 UT and lasted for 59 ks.
The detailed method of EPIC-MOS data reduction, and the timing, spectroscopic and flare analysis are given in Pandey \& Singh (2008). Here, we used the 0.3-10.0
keV X-ray data observed from EPIC-MOS1.  The X-ray light curve of $\xi$ Boo is shown in  Figure \ref{fig:lc},
where flare regions are represented by F1 and F2, and the quiescent state is represented by Q. The light curve was binned with 40s. In the X-ray light curve of $\xi$ Boo,  a period of  heightened coronal emission 'U' was identified after the flare F2, where the average flux was 1.8 times more than the quiescent state. We have chosen the U region for our wavelet and periodogram analyses because of the stable evolution of the integrated 0.3-10 keV light curve after the flare F2.
To study the emission
in the  U regions only, we fitted the spectral data using 1-temperature plasma model apec
(Smith et al. 2001), with the quiescent emission taken into account by including its best-fit
2-temperature model as a frozen background contribution. The best fit temperature and emission measure during
the U part of the light curve are $8.7\pm0.3$ MK and $3.8\pm0.5\times10^{51}$ cm$^{-3}$, respectively
with reduced $\chi^2$ per degree of freedom of 0.99. During the flare F2, the average values of
temperature and emission measures were found to be $12.0\pm0.6$ MK and $3.5\pm0.1\times 10^{51}$ cm$^{-3}$,
respectively.

Line fluxes and positions were measured using the XSPEC package
by fitting simultaneously the RGS1 and RGS2 spectra with a
sum of narrow Gaussian emission lines convolved with the response matrices of the RGS instruments. The continuum emission was described using Bremstrahlung models at the temperatures of the plasma components inferred from the analysis of EPIC-MOS1 data. The emission measure derived from the analysis of the EPIC data  was used to freeze the continuum normalization. For the present purpose we give the line fluxes of He-like triplets from {\sc O vii} in Table 1.

\begin{deluxetable}{cccc}
\tablecolumns{4}
\tablewidth{-0pt}
\tablecaption{\leftskip0mm{Observed {\sc O vii} line fluxes during the Q and F2+U states}}
\label{tab:flux}
\tablehead{
\multicolumn{2}{c}{Quiescent (Q)}&\multicolumn{2}{c}{Flaring (F2+U)}\\
$\lambda (\AA)$ & Flux$^a$  & $\lambda (\AA)$ & Flux$^a$  }
\startdata
22.08(f)&$2.1\pm0.2$ & 22.08(f)&$2.2\pm0.3$\\
21.78(i)&$0.8\pm0.2$ & 21.77(i)&$0.9\pm0.2$\\
21.58(r)&$3.0\pm0.3$ & 21.57(r)&$3.4\pm0.3$\\
&&&\\
\enddata
\tablenotetext{~}{$^a$  Measured flux in $10^{-4}$ photons/cm$^2$/s}
\vspace{-0.5cm}
\end{deluxetable}

\section{Analysis and Results}
\subsection{Density Measurement}
The He-like transitions, consisting of the resonance line (r) 1s$^2$ ~$^1$S$_0$ - 1s2p $^1$p$_1$, the intercombination line (i)  1s$^2$ $^1$S$_0$ - 1s.2p $^3$P1, and the forbidden line (f) 1s$^2$ $^1$S$_0$ - 1s.2s $^3$S$_1$ are density- and temperature-dependent (e.g. Gabriel \& Jordan 1969). The intensity ratio G = (i+f)/r varies with temperature and the ratio R =f/i varies with electron density due to collisional coupling between the metastable $2^3$S upper level of forbidden line and the $2^3$P upper level of the intercombination line. In the RGS wavelength, the {\sc O vii} lines are clean, resolved and potentially suited to diagnose electron density and temperature (see Fig. 2). These lines at 21.60\AA (r), 21.80\AA(i) and 22.10\AA(f) have been used to obtain the temperature and density values from the G- and R-ratio using CHIANTI database (version 5.2.1; Landi et al. 2006). The G-ratio for $\xi$ Boo implies a temperature of $1.8\pm0.6$ and $2.0\pm0.6$ MK during the quiescent and flare states, respectively. For these temperatures, we derive electron densities of $n_e = 1.3_{-0.5}^{+1.2} \times 10^{10}$ cm$^{-3}$ for the quiescent state (Q) and $n_o = 1.6_{-0.8}^{+1.2}\times10^{10}$ cm$^{-3}$ for flare state (F2+U). The electron density was slightly higher during the flare state. However, this  increase is  well within the  $1 \sigma$ level.

\begin{figure}
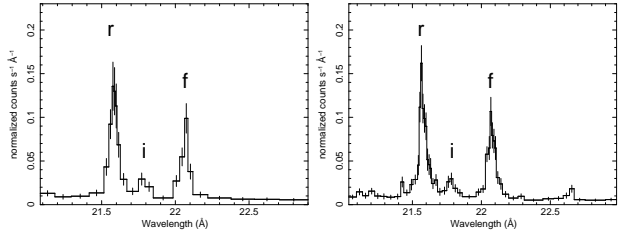

\begin{center}
\includegraphics[width=3cm,height=4cm,angle=-90]{Q_rif.ps}
\includegraphics[width=3cm,height=4cm,angle=-90]{F2U_rif.ps}
\caption{The {\sc O vii} line of $\xi$ Boo observed by RGS1.  Left panel: Quiescent state 'Q' and right panel: Flaring state 'F2+U'.}
\end{center}
\label{fig:rif}
\end{figure}

\subsection{Loop Length}
\label{sec:looplength}
The length of the stellar loops associated with the observed flaring activity can be estimated from the various loop models. Below we derive loop length  from four different models.

\subsubsection{Hydrodynamic Flare model}
The details of this method are given in Pandey \& Singh (2008). The derived loop length for the flare F2 is found to be of  $7.9\pm 6 \times 10^{10}$ cm. However, the value of $\zeta$ (=2; slope of density temperature diagram) is outside the domain of the validity of the method. Therefore, the estimated loop length from this method may not be actual.
Alternatively, Reale (2007)
derived the loop length ($L$) from the rise phase and peak phase of the flare.
The loop length of the flare F2 from this approach is estimated to be $3.6\pm0.8 \times 10^{10}$ cm.

\subsubsection{Pressure Balance method}
Shibata \& Yokoyama (2002) assume that the gas pressure is balanced by the
magnetic pressure and give the equation for the magnetic field ($B$) and the loop length($L$):

{\scriptsize
\begin{equation}
B = 50 \left( \frac{EM}{10^{48} cm^{-3}}\right)^{-1/5} \left( \frac{n_e}{10^9 cm^{-3}}\right)^{3/10}\left(\frac{T}{10^7K}\right)^{17/10}
\label{eq:B}
\end{equation}

\begin{equation}
L = 10^9 \left( \frac{EM}{10^{48} cm^{-3}}\right)^{3/5} \left( \frac{n_e}{10^9 cm^{-3}}\right)^{-2/5}\left(\frac{T}{10^7K}\right)^{-8/5}
\label{eq:L}
\end{equation}
}
\noindent
where $n_e$ is the density outside the flaring loop.
By applying the average values of EM  and T  during the flare, we obtained a magnetic field strength of $29.4\pm6.0$ G and a loop length of $3.6\pm0.9 \times 10^{10}$ cm for the flare F2.

\subsubsection{Rise and decay time method}
The approach is described in Hawley et al. (1995). In this approach, the loop length  has been derived by solving the time-dependent energy equation in rise and decay phase separately in different limits: strong evaporation driven by heating in the rise phase and strong condensation driven by radiation in the decay phase. The time rate change of the loop pressure ({\it \.{p}}) for the spatial average is given by

\begin{equation}
\frac{3}{2} {\dot{p}} =  Q- R
\end{equation}

\noindent
where $Q$ is the volumetric flaring heating rate, and $R$ is the optically thin cooling rate.
The rise phase is dominated by strong evaporative heating, i.e., $Q >> R$, while the decay phase is dominated by cooling and strong condensation, i.e., $R>>Q$. At the loop top, there is an equilibrium, i.e., $Q = R$ and the loop length  can be given as
{\scriptsize
\begin{equation}
L = \frac{1500}{(1-x_d^{1.58})^{4/7}}.\tau_d^{4/7}.\tau_r^{3/7}.T_A^{1/2}
\end{equation}
}

\noindent
where $\tau_d$ is the decay time, $\tau_r$ is the rise time, $T_A$ is the temperature at flare apex, and $x_d = c_q/c_{max}$. $c_{max}$ (=3.16 counts/s) and $c_{q}$ (=1.24 counts/s) are count rate at the peak of the flare and the count  rate during the quiescent state, respectively.  Using this approach, the loop length for the flare F2 is estimated to be $3.9\pm0.5\times10^{10}$ cm. The values of $\tau_d$(=9885 s), $\tau_r$ (= 3351 s), and $T_A$ (=13.3 MK) are taken from Pandey \& Singh (2008).

\subsubsection{Haisch's Approach}
     Given an estimate of two measured quantities, the emission measure (EM) and the decay time scale of the flare ($\tau_d$ ) Haisch's approach (Haisch 1983) leads to the following expression for the loop length ($L$)

{\scriptsize
\begin{eqnarray}
L = 5\times10^{-6} EM^{1/4}\tau_d^{3/4}
\end{eqnarray}
}

\noindent
Using this approach, we obtained a loop length of $3.8\pm0.2\times10^{10}$ cm.
The minimum magnetic field strength is estimated from the condition
$B^2/8\pi \leq 2 n_o k_B T$, where $k_B$ is Boltzmann's constant.
The minimum  magnetic field strength thus obtained was 36 Gauss.
This lies in the upper limit of the magnetic field determined by the pressure balance method.

The loop length derived from the  above all four methods is consistent with each other.

\subsection{Wavelet and Periodogram Analyses}
\label{wavelet}
We have used the wavelet analysis IDL code ``Randomlet"
developed by E. O'Shea.
The program executes a randomization test (Linnell Nemec \& Nemec 1985; O'Shea et al. 2001)
which is an additional feature along with the standard wavelet analysis code (Torrence \& Compo 1998) to examine the existence of statistically significant real periodicities in the time series data.  The advantage of using this test is that it is distribution free or non parametric, i.e., it is not constrained by any specific noise model such as Poisson, Gaussian, etc. Using this technique, many important results have been published by analyzing approximately evenly sampled data (e.g., Banerjee et al. 2001; O'Shea et al. 2001, 2007;  Srivastava et al. 2008a,b; O'Shea \& Doyle 2009; Gupta et al. 2009).

\begin{figure}
\begin{center}
\subfigure[]{\includegraphics[width=70mm,angle=90]{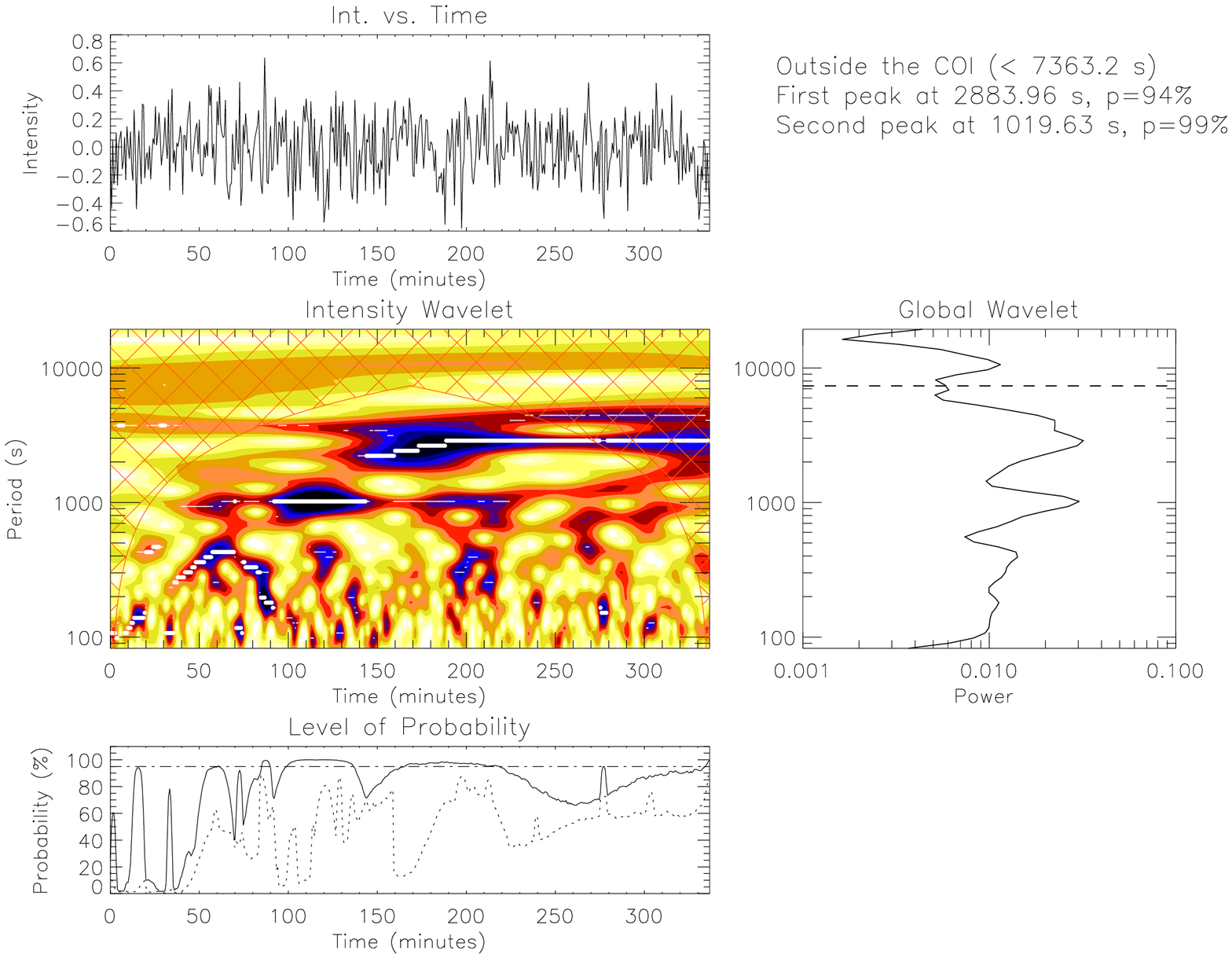}}
\subfigure[]{\includegraphics[width=50mm,angle=-90]{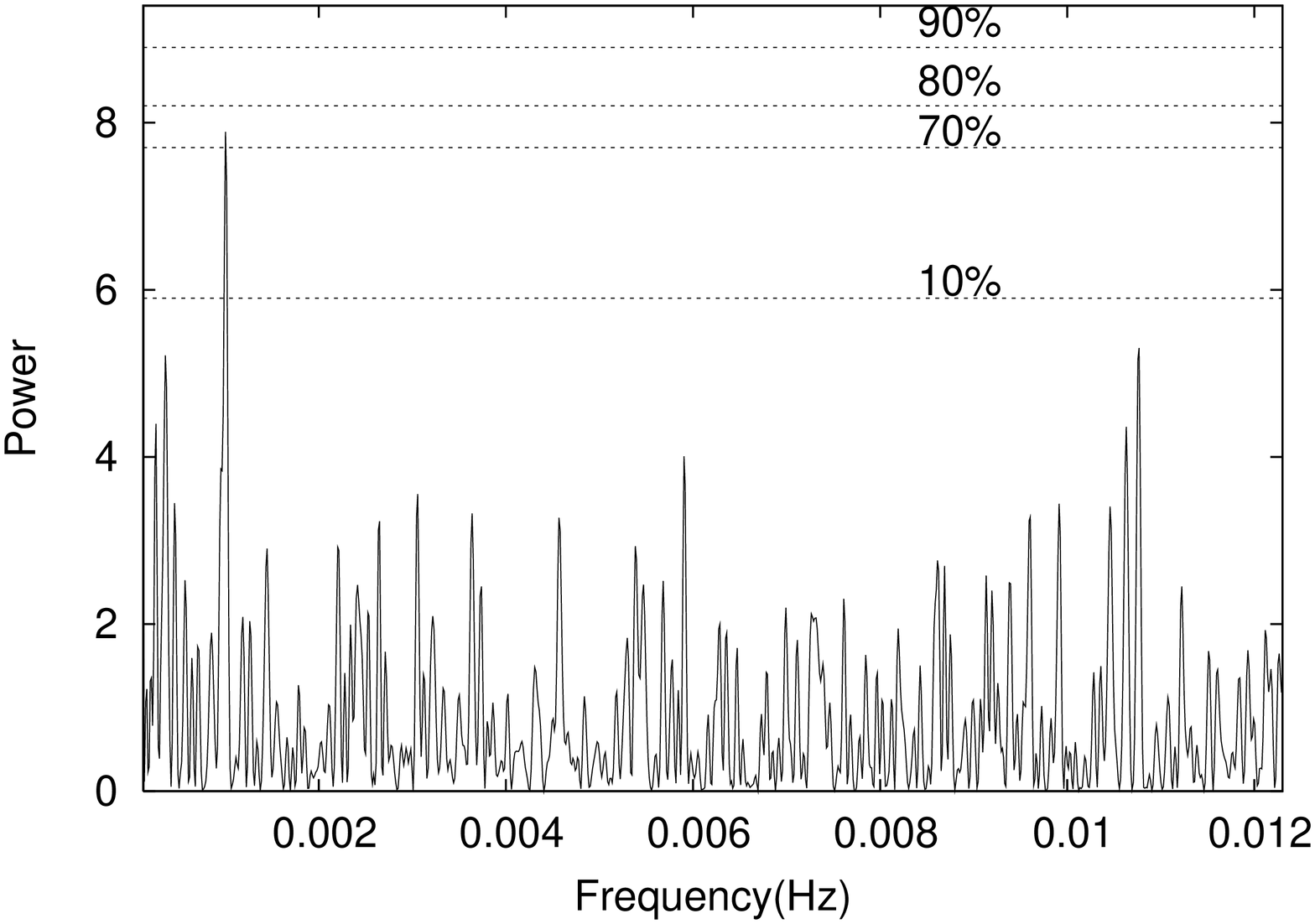}}
\caption{(a) The wavelet result for U part of the X-ray light curve of $\xi$ Boo. The top panel shows the variation of intensity, the wavelet power spectrum is given in the middle panel, and the probability in the bottom panel. (b) Power spectrum from periodogram analysis.}
\end{center}
\label{fig:period}
\end{figure}

The wavelet power transforms of the U part of the X-ray light curve of cadence 40 s (see the top-left panel of Fig. 3a) are shown in the middle left panel of Fig. 3a, where  the darkest regions show the most enhanced oscillatory power in the intensity wavelet spectrum. The cross-hatched areas are the cone of influence (COI), the region of the power spectrum where edge effects, due to the finite lengths of the time series, are likely to dominate.
The maximum allowed period from COI, where the edge effect is more effective, is $\sim 7363$s.
In our wavelet analysis, we only consider the power peaks and corresponding real periods (i.e. probability $>$ 95\%) below this threshold.
The period with maximum power detectable outside the COI is
$\sim$2883 s with a probability of $\sim$94\% and the repetition of only $\sim3$
cycles in the time series. However, another peak is visible
at a period of $\sim$1019 s with a probability of $\sim$99\% and the
repetition of $\sim$11 cycles. Therefore, only the periodicity of
$\sim$1019 s satisfies the recently reported
 more strict  criteria of O'Shea \& Doyle (2009), i.e. at least four cycle of repetition over the life time of the oscillations.  The middle-right panel in  Fig. 3a shows the global wavelet power spectra of the time series from which the statistically significant period ($\sim$1019s) is selected.
In Fig. 3a, the bottom panel shows the probabilities of the presence of two specific frequencies (or periods) corresponding to the first and second highest peaks shown in the middle-left panel as functions of the time after the start of the observations . The solid line represents the probability corresponding to the major power peak of time series data, while the dotted line corresponds to the secondary power peak.
Wavelet analysis  was also performed in the decay phase of the flare F2, and no significant periodicity was found.

We have also performed the periodogram analysis (Scargle 1982) in the U part of the X-ray light curve (see Fig. 3b).  In the power spectrum, the highest peak corresponds to a period of $\sim$1005 s (probability $\sim 75$\%), which is consistent with that obtained from wavelet analysis.
Therefore, only the period of $\sim$ 1019 s appears to be statistically significant in our analyses.

\begin{deluxetable*}{llcccc}
\tablecolumns{6}
\tablewidth{-0pt}
\tabletypesize{\scriptsize}
\tablecaption{\leftskip0mm{Derived parameters}}
\label{tab:comp}
\tablehead{
Model & Loop Length & \multicolumn{3}{c}{Theoretically estimated Period} & Observationally \\
\cline{3-5}
      & ($10^{10}$) cm& Slow Mode  & Fast-Kink mode$^{ab}$ & Fast Sausage Mode$^b$ & Estimated \\
      &              &   (s)      &     (s)        &    (s)   &period (s)}
\startdata
Hydrodynamic      &$3.6\pm0.8$&$1586\pm353$  &$1004\pm391$  &$313\pm121$&\\
Rise and decay    &$3.9\pm0.5$&$1718\pm222$  &$1087\pm374$  &$339\pm116$&\\
Pressure balance  &$3.6\pm0.9$&$1586\pm377$  &$1004\pm407$  &$313\pm127$&1019\\
Haisch's approach &$3.8\pm0.2$&$1674\pm92~$  &$1059\pm343$  &$330\pm107$&\\
&&&&&\\
\enddata
\tablenotetext{~}{$^a$ Large error bars are due to the large density error, $^b$ Loop width was determined by assuming $a/L = 0.1$}
\vspace{-0.5cm}
\end{deluxetable*}

\subsection{Fast-kink Oscillations}
The observed periodicity may be the likely signature of MHD oscillations associated with the stellar flare loops.
The overdense magnetic loops are pressure balanced structures and may contain fast-kink and sausage oscillation modes
with phase speed greater than the local Alfv\'enic speed ($\omega/k > v_{A}$), and slow acoustic modes
with sub-Alfv\'enic speed ($c_{0} < v_{A}$) under coronal conditions.

The expressions for the oscillation periods of slow ($P_{s}$), fast-kink ($P_{fk}$), and fast-sausage ($P_{fs}$) 
modes are given as follows (Edwin \& Roberts, 1983; Roberts et al., 1984; Aschwanden et al., 1999) :
{\scriptsize
\begin{equation}
P_{s} = \frac{2 L}{jc_{T}}=\frac{2 L}{j c_{o}}\left[1 + \left(\frac{c_{o}}{v_{A}}\right)^{2}\right]^{1/2}
\approx 1300 \frac{L_{10}} {\sqrt T_{6}} s
\end{equation}

\begin{equation}
P_{fk} = \frac{2L}{jc_{k}} = 4\pi^{1/2}\frac{L}{j}\left( \frac{\rho_{o} + \rho_{e}}{B_{o}^{2} + B_{e}^{2}}\right)^{1/2}
\approx 205 \frac{L_{10} \sqrt{n_{9}}}{B_{10}} s
\end{equation}

\begin{equation}
P_{fs} = \frac{2\pi a}{c_{k}} = 4\pi^{3/2}a\left( \frac{\rho_{o} + \rho_{e}}{B_{o}^{2} + B_{e}^{2}}\right)^{1/2}
\approx 6.4 \frac{a_{8} \sqrt{n_{9}}}{B_{10}} s
\end{equation}
}
\noindent
where $L$, $a$, $\rho$, and $B$ are the loop length, the loop width, the ion mass density and magnetic field,
respectively. The $c_{T} = {c_{o} v_{A}}/ {(c_{o}^2+v_{A}^2)}^{1/2}$, where $c_{o}$ and $v_{A}$
are the sound  and Alfv\'enic speeds respectively. The $c_{k}$ is the phase speed. Under coronal conditions, $B_{o} \approx B_{e}$,
$\rho_{e} << \rho_{o}$, $ n_H \approx n_e$, and $c_{o} << v_{A}$. The subscripts $'o'$ and $'e'$
stands for 'inside', and 'outside' of the loop. The $j$ is the number of nodes in fast-kink and slow oscillations,
and is equal to 1 for the fundamental mode. The approximate relation in Equations 6-8 are given in coronal conditions
for the fundamental oscillation periods of different modes in magnetic loops. Therefore, these approximate
equations are also valid for the stellar loops in the coronae of magnetically active Sun-like stars.
The parameters are expressed in units of $L_{10} = L/10^{10} cm $, $a_{8} = a/10^{8} cm$, $n_{9} = n_o/10^{9} cm^{-3}$, $T = T_{e}/10^{6} K$, and $B_{10} = B/10 G$. The loop width was determined by assuming $a = 0.1 L$ (Shimizu 1995) for a single loop.

Using the observationally estimated loop length and width, density, magnetic field,
and Equations (6)-(8), the estimated  periods for different oscillation modes are given in Table 2.
The observationally derived oscillation period $\sim1019$  s approximately
matches with the theoretically derived oscillation period  for a fundamental fast kink mode.

\section{Discussion and Conclusions \label{sec:disc}}
We found the first signature of fundamental kink oscillations with period of $\sim 1019$ s in the magnetically active star $\xi$ Boo observed by EPIC-MOS of {\it XMM-Newton}, which may be formed  either by the superposition of oppositely propagating fast kink waves or due to flare generated disturbances near the loop apex. The flaring activity may be the possible mechanism of the generation of such oscillations in the stellar loops. Recently, Cooper et al. (2003) have found that the transverse kink modes of fast magneto-acoustic
waves may modulate the emission observable with imaging telescopes despite their incompressible nature. This is possible only when the axis of an oscillating loop is not orientated perpendicular to the observer's line of sight.

Resonant absorption is the most efficient mechanism
theorized for the damping of a kink mode in which energy of the mode is transferred to the localized Alfv\'{e}nic oscillations of the inhomogeneous layers at the loop boundary (Lee \& Roberts 1986). Intensity oscillations  due to the fast-kink mode of magneto-acoustic waves do not show any decay of the amplitude during the initial 250 minutes in the post-flare phase observations of $\xi$ Boo. This is probably due to an impulsive driver, which forces such kind of oscillations and dominates over the dissipative processes during  the flare F2 of $\xi$ Boo.

In conclusion, we report the first observational evidence of fundamental fast-kink oscillations in stellar loops during a post-flare phase of heightened emission in the cool active star $\xi$ Boo.
 These oscillations are a unique tool for the estimation of several key parameters of coronae, which are not directly measurable, e.g., the magnetic field. Some of the seismological techniques developed for the solar corona, e.g., based upon the ratio of the periods of kink modes, do not require the spatial resolution, and hence can be applied to estimate the density scale heights of the stellar coronae. Therefore, based on the solar analogy, such observations may shed new lights on the dynamics of the magnetically structured stellar coronae and its local plasma conditions. However, a future  multiwavelength observational search is required to investigate various kinds of MHD modes to probe the local conditions of plasma in magnetically active stars.

\vspace{-0.5cm}

\acknowledgments
We thank referee for his valuable comments and suggestions.  We thank E. O'Shea for 'Randomlet'. We also thank  M. J. Aschwanden, V. M. Nakariakov, L. Ofman,  R. Sagar, and  T. V. Zaqarashvili for carefully reading the manuscript and valuable suggestions. The research is based on observations obtained with {\it XMM-Newton}, an ESA science mission with instruments and contributions directly funded by ESA Member States and NASA.

\end{document}